\documentclass[reprint,
superscriptaddress,
amsmath,amssymb,
aps,
prl,
floatfix,
]{revtex4-2}

\usepackage{graphicx}
\usepackage{dcolumn}
\usepackage{bm}
\usepackage{hyperref}
\usepackage[usenames,dvipsnames]{color}

\begin{document}

\preprint{APS/123-QED}

\title{Edge non-collinear magnetism in nanoribbons of Fe$_3$GeTe$_2$ and Fe$_3$GaTe$_2$}

\author{Ramon Cardias}
\email{ramon_cardias@hotmail.com}
\affiliation{ 
Instituto de Física, Universidade Federal Fluminense, 24210-346, Niterói RJ, Brazil
}

\author{Anders Bergman}
\affiliation{
Department of Physics and Astronomy, Uppsala University, Box 516,
SE-75120 Uppsala, Sweden
}

\author{Hugo U. R. Strand}
\affiliation{
School of Science and Technology, \"Orebro University, SE-70182 Örebro,
Sweden
}

\author{R. B. Muniz}
\affiliation{
Instituto de Física, Universidade Federal Fluminense, 24210-346, Niterói RJ, Brazil
}

\author{Marcio Costa}
\affiliation{
Instituto de Física, Universidade Federal Fluminense, 24210-346, Niterói RJ, Brazil
}

\date{\today}

\begin{abstract}

Fe$_3$GeTe$_2$ and Fe$_3$GaTe$_2$ are ferromagnetic conducting materials of van der Waals-type with unique magnetic properties that are highly promising for the development of new spintronic, orbitronic and magnonic devices. Even in the form of two-dimensional-like ultrathin films, they exhibit relatively high Curie temperature, magnetic anisotropy perpendicular to the atomic planes and multiple types of Hall effects. We explore nanoribbons made from single layers of these materials and show that they display non-collinear magnetic ordering at their edges. This magnetic inhomogeneity allows angular momentum currents to generate magnetic torques at the sample edges, regardless of their polarization direction, significantly enhancing the effectiveness of magnetization manipulation in these systems.
We also demonstrate that it is possible to rapidly reverse the magnetization direction of these nanostructures by means of spin-orbit and spin-transfer torques with rather low current densities, making them quite propitious for non-volatile magnetic memory units. 

\end{abstract}

\maketitle

\textit{Introduction - }The ability to manipulate the magnetic ordering of nanoscale systems is of great interest to the magnetic industry.
It enables the design and production of smaller and more efficient devices to process and store information. Spintronics, magnonics, and orbitronics have progressed greatly in recent years and are offering ingenious and innovative mechanisms to accomplish this, utilizing the transport of charge, spin, and orbital angular momentum across nanostructures. \cite{sinova_spin_2015,jo_spintronics_2024, jo_spintronics_2024, ralph_spin_2008, hals_phenomenology_2013} 

Two-dimensional (2D)-like materials, such as ultrathin films and multilayers, are currently receiving considerable attention due to their versatilities, particularly the magnetic van der Waals (vdW) ones \cite{wang_magnetic_2022}, 
which can be exfoliated into ultrathin films and deposited with relative ease on different substrates, forming heterostructures with more comprehensive properties and functionalities. Among them, Fe$_3$GeTe$_2$ (FGeT) stands out as the first synthesized 2D metallic ferromagnet \cite{deng_gate-tunable_2018, fei_two-dimensional_2018}, and is a very promising material for next-generation spintronics devices. It exhibits out-of-plane magnetic anisotropy, sizeable spin-orbit interaction, and a relatively high anomalous Hall effect (AHE) \cite{wang_anisotropic_2017,  kim_large_2018}. Its Curie temperature ($T_c$) may be controlled by doping \cite{jang_origin_2020}, ionic gating \cite{deng_gate-tunable_2018}, or through interaction with suitable substrates. 

More recently, Fe$_3$GaTe$_2$ (FGaT) has been also synthesized with Ga atoms replacing Ge. The magnetic properties of the two materials are similar, but Fe$_3$GaTe$_2$ is even more attractive for practical use because its $T_c$ value is above room temperature for thick films, and remains rather large ($\approx 240$K) for the monolayer \cite{zhang_above-room-temperature_2022, wang_hard_2024}.

Different ways of manipulating the magnetic properties of nano-structured systems have been extensively investigated lately using current sources of distinct natures \cite {miron_perpendicular_2011, zhang_gigantic_2021, zhang_highly_2021, wang_room_2023,  yan_highly_2024, alghamdi_highly_2019, kajale_current-induced_2024, dai_interfacial_2024, zhou_controllable_2021}. A particularly promising strategy involves using the flow of angular momentum with a specific polarization to apply torque to local magnetic moments and excite spin dynamics within the magnetic system.
Spin-polarized electric currents, as well as pure spin- and orbital-angular momentum currents have been employed for such purpose. Spin-polarized electrical currents are typically generated by using magnetic fields or ferromagnetic (FM) materials as polarizers, while pure spin currents are most commonly produced by spin pumping or the spin Hall effect (SHE). These currents are usually generated in auxiliary materials and then injected to manipulate the magnetization of adjacent magnetic units. The effectiveness of the torque they produce depends on both the direction and degree of spin polarization. Notably, both spin-transfer torque (STT) and the spin-orbit torque (SOT) are zero if the spin-current polarization is parallel to the local magnetic moment. 

Recently, significant investigations have been made into the production of orbital angular momentum currents and their role in exciting spin dynamics. Although generating these currents through the orbital Hall effect does not require spin-orbit coupling (SOC) in the host material, the orbital torque (OT) depends on spin-orbit interaction (SOI) within the magnetic unit. Both FGeT and FGaT exhibit sufficiently large SOC to generate SOT and OT, which may be used to control their magnetization.    
 
In addition to the relatively high anomalous Hall effect AHE observed in these materials \cite{lin_layer-dependent_2019, zhang_above-room-temperature_2022}, an electric current flowing through them may also give rise to the conventional (SHE) and the spin anomalous Hall effect (sAHE) \cite{iihama_spin-transfer_2018} due to SOI. More recently, additional anomalous contributions to spin currents induced by charge currents in ferro- and anti-ferromagnets have been predicted \cite{wang_anomalous_2021, salemi_theory_2022}, widening the prospects for magnetization manipulations, although some still wait experimental confirmation. 

Both systems crystallize in a hexagonal structure, belonging to the crystallographic point group $D_{6h}$ of the space group P6$_3$/mmc (No. 194). It consists of layers stacked along the c-axis, each of which comprising a Fe$_3$Ge(Ga) sublayer sandwiched between two Te atomic planes. A single layer of Fe$_3$Ge(Ga)Te$_2$ contains five atomic planes, and belongs to the point group symmetry $D_{3h}$. Fig 1(a) illustrates its lattice structure and unit cell, highlighting the two nonequivalent Fe atoms that occupy octahedral (Fe$_\text{I}$) and tetrahedral (Fe$_\text{II}$) sites, respectively.

Here, we investigate nanoribbons made from single layers of FGeT and FGaT with armchair-shaped edges. The presence of these borders reduces the $D_{3h}$ symmetry of the monolayer, giving rise to a complex competition between the bilinear exchange and the Dzyaloshinskii-Moriya (DM) interactions, producing a ground state with noncollinear magnetic ordering at the edges. This feature facilitates the manipulation of magnetization of these systems through angular momentum torques, as the edges experience torque regardless of the angular momentum polarization direction.

 We shall start with density functional theory (DFT) band structure calculations for single layers of FGeT and FGaT, using the plane-wave-based code \textsc{Quantum ESPRESSO}. The results are shown in Supplementary Material (SM) alongside our band calculations using pseudo-atomic orbitals with the \textsf{PAOFLOW} code. Clearly, the two sets are in excellent agreement. Both materials exhibit FM ground state with perpendicular magnetic anisotropy. The calculated values for the Fe magnetic moments and for the single ion magnetic anisotropy energy (MAE) are also listed in the SM.

Fig.~S1 of the SM also displays the anomalous Hall conductivity $\sigma^{0}_{xy}$, together with the spin Hall $\sigma^{z(\text{S})}_{xy}$ and orbital Hall $\sigma^{z(\text{O})}_{xy}$ conductivities with polarization $\hat{z}$ perpendicular to the layers, calculated as functions of energy for FGeT and FGaT. They were computed using the PAO Hamiltonian and linear response theory, as detailed in the SM. Notably, the value of $\sigma^{z(\text{O})}_{xy}$ around the Fermi energy (E$_\text{F}$) is a factor 8.5(7.2) larger than $\sigma^{z(\text{S})}_{xy}$ for FeGeT(FeGaT), indicating that orbital torques may be also useful for magnetization manipulation in these systems.

\begin{figure}[htp]
    \centering
            \includegraphics[width=0.9\columnwidth]{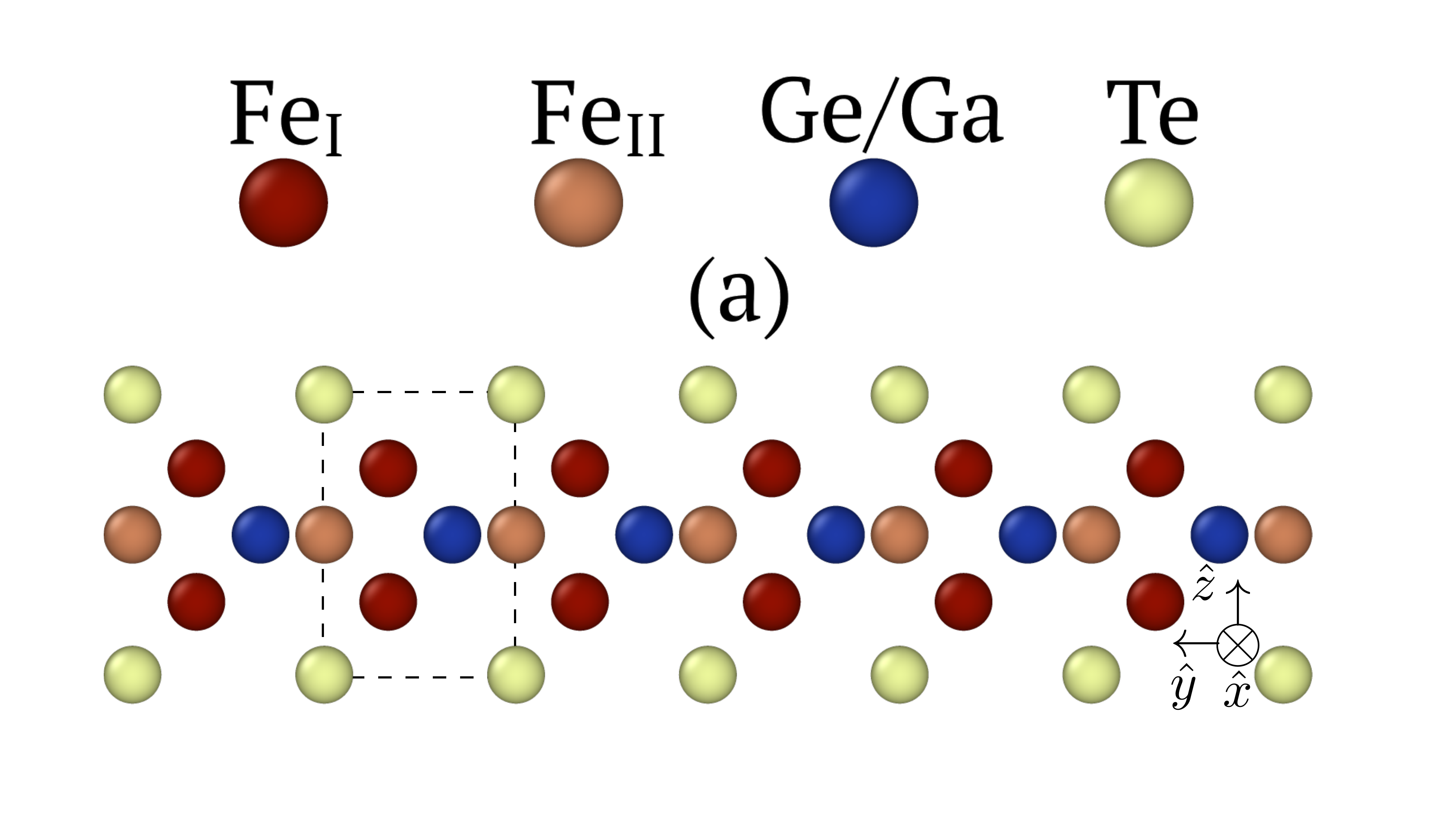} \\
            \vspace{0.5cm} 
            \includegraphics[width=0.9\columnwidth]{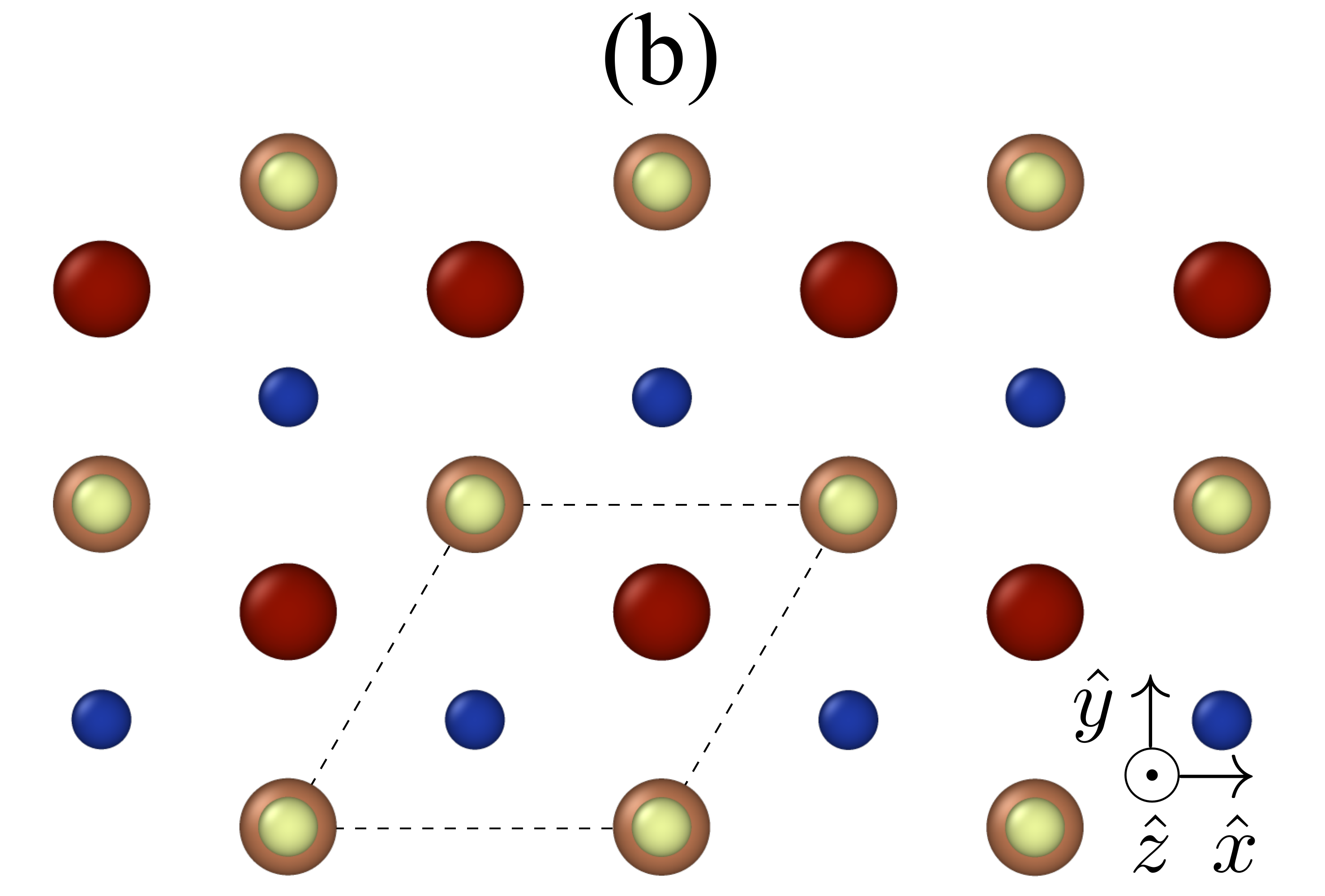}
    \caption{Schematic representation of the crystalline structure of a monolayer of Fe$_3$Ge(Ga)Te$_2$, side view (a), top view (b). The dashed lines portray the unit cell boundaries.}
    \label{fig:fgtsnanoribbon}
\end{figure}

The energy bands calculated using DFT for nanoribbons with armchair edges extracted from monolayes of Fe$_3$GeTe$_2$ and Fe$_3$GaTe$_2$ are shown in Fig.~S2 of SM. The color coding scheme illustrates the spatial character of the corresponding eigenstates, where red denotes high probability amplitudes at the edge sites and blue represents high probability amplitudes in the central region of the ribbon. Both ribbons are approximately 18\AA~in breadth.

\begin{figure}[htp]
    \centering
    \includegraphics[width=1.0\columnwidth]{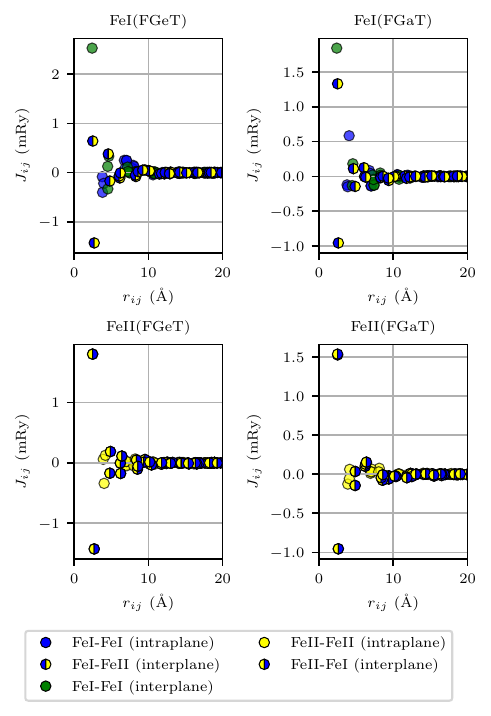}
    \caption{Effective exchange couplings between the magnetic moment of a Fe atom (Fe$_\text{I}$ - top panels or Fe$_\text{II}$ - bottom panels) located at the edge of an armchair nanoribbon and its neighboring Fe atoms, calculated as functions of their interatomic distances. Solid circles represent interactions between Fe atoms of the same type: (blue) Fe$_\text{I}$-Fe$_\text{I}$ in-plane, (green) Fe$_\text{I}$-Fe$_\text{I}$ inter-plane, (yellow) Fe$_\text{II}$-Fe$_\text{II}$ in-plane. Inter-plane interactions between Fe atoms of distinct types are depicted by circles with two colors: (blue left/ yellow right) Fe$_\text{I}$-Fe$_\text{II}$ and (yellow left/ blue right) Fe$_\text{II}$-Fe$_\text{I}$. }
    \label{fig:magneticpar}
\end{figure}

An useful strategy for exploring the ground state spin configuration of these magnetic systems is to represent them by an effective spin Hamiltonian as follows: 
\begin{equation}
H = -\sum_{i,j} J_{ij} \hat{m}_i \hat{m}_j - \sum_{i,j} \vec{D}_{ij} \cdot \left(\hat{m}_i \times \hat{m}_j\right) - \mathcal{K} \sum_{i} \left(\hat{m}_i \cdot \hat{m}_i^k \right)^2~\,
\label{eq:Hamilton0}
\end{equation}
where $J_{ij}$ represents the pairwise effective exchange interaction between local magnetic moments $\vec{m}_i = m_i \hat{m}_i$ and $\vec{m}_j = m_j \hat{m}_j$ located at sites $i$ and $j$, respectively. Here, $\hat{m}_i$ represents the unit vector along the local moment direction and $m_i$ its magnitude. $\vec{D}_{ij}$ denotes the DM vector, associated with the effective DM interaction due to SOC. The last term typifies the single-ion anisotropy with intensity proportional to $\mathcal{K}$, and $\hat{m}_i^k$ indicates the unit vector along the direction of the uniaxial anisotropy. The magnitudes of the moments have been incorporated into the interaction constants, which are expressed in energy units. 

The effective spin interactions $J_{ij}$, $\vec{D}_{ij}$ and $\mathcal{K}$ can be completely determined from electronic structure calculations, as briefly described in sections S2 and S3 of the SM. Here, we use the Liechtenstein, Katsnelson, Antropov, and Gubanov (LKAG) and Green's function formalisms \cite{szilva_quantitative_2023,liechtenstein_local_1987, cardias_first-principles_2020, frota-pessoa_exchange_2000} to compute $J_{ij}$ and $\vec{D }_{ij}$ for any pair of Fe magnetic moments. The Green's functions involved are expanded in terms of Chebyshev polynomials using the kernel polynomial method \cite{weise_kernel_2006}.

Fig.\ref{fig:fgtsnanoribbon} shows that the Fe atoms in a single layer of FeGe(Ga)T are arranged in three atomic planes. The one in the middle of the layer contains only Fe$_\text{II}$ and Ge(Ga) atoms, and is sandwiched by two others, formed solely by Fe$_\text{I}$ atoms. It is instructive to calculate the values of $J_{ij}$ and $\vec{D}_{ij}$ for the single layer as a reference for comparison with those of the nanoribbons. The results are displayed in Fig.~S3 as functions of the interatomic distances, highlighting the inter- and in-plane pairs of Fe atoms. 

It is worth mentioning that the minimization of the total energy, using Eq. \ref{eq:Hamilton0} along with the values of $J_{ij}$ and $\mathcal{K}$ for the single layer confirms the FM configuration of the ground state with magnetic moments oriented perpendicular to the layer, consistent with DFT calculations \footnote{The contribution to the total energy coming from the Dzyaloshinskii-Moriya interaction vanishes in the single layer due to symmetry.}.

The same formalism has been employed to calculate the values of $J_{ij}$ and $\vec{D}_{ij}$ for FGeT and FGaT nanoribbons with armchair borders.
Far from the edges, the results are very similar to those for the single layer, as expected. However, near them, the differences are significant. Hence, to simplify the presentation, we have selected a Fe atom located at one of the edges and illustrated its interaction with neighboring Fe atoms in Fig.~\ref{fig:magneticpar} \footnote{Here, the ends of the nanoribbon are equivalent due to the mirror symmetry $\mathcal{M}_x$ or $\tau\mathcal{M}_x$ of the armchair-edged ribbons}.
We note that the highest value of $J_{ij}$ is the interplane FM coupling between the nearest-neighbor Fe$_\text{I}$ atoms, as Fig. \ref{fig:magneticpar} illustrates. Next come the interplane couplings between Fe$_\text{I}$ and its nearest-neighbor Fe$_\text{II}$ atoms, which exhibit both FM and antiferromagnetic (AFM) contributions, Then, the in-plane couplings between next-nearest neighbors Fe$_\text{I}$ atoms, which are AFM in the case of FGeT, but display both FM and AFM contributions in FGaT.
Slight variations in the interatomic distances between originally equidistant Fe atoms may occur because of spatial relaxations considered in our DFT calculations. The $J_{ij}$ values between these nearly equidistant Fe atoms can differ significantly, as they become nonequivalent due to the breaking of translation symmetry along the transverse direction of the nanoribbon.
The significant changes in $\vec{D}_{ij}$ caused by the loss of translational symmetry in the transverse direction of the nanoribbons are illustrated in Figure~S4.

\begin{figure}[htp]
    \centering
    \begin{tabular}{c}
        \begin{tabular}{c}
            (a) \\ \includegraphics[width=0.95\columnwidth]{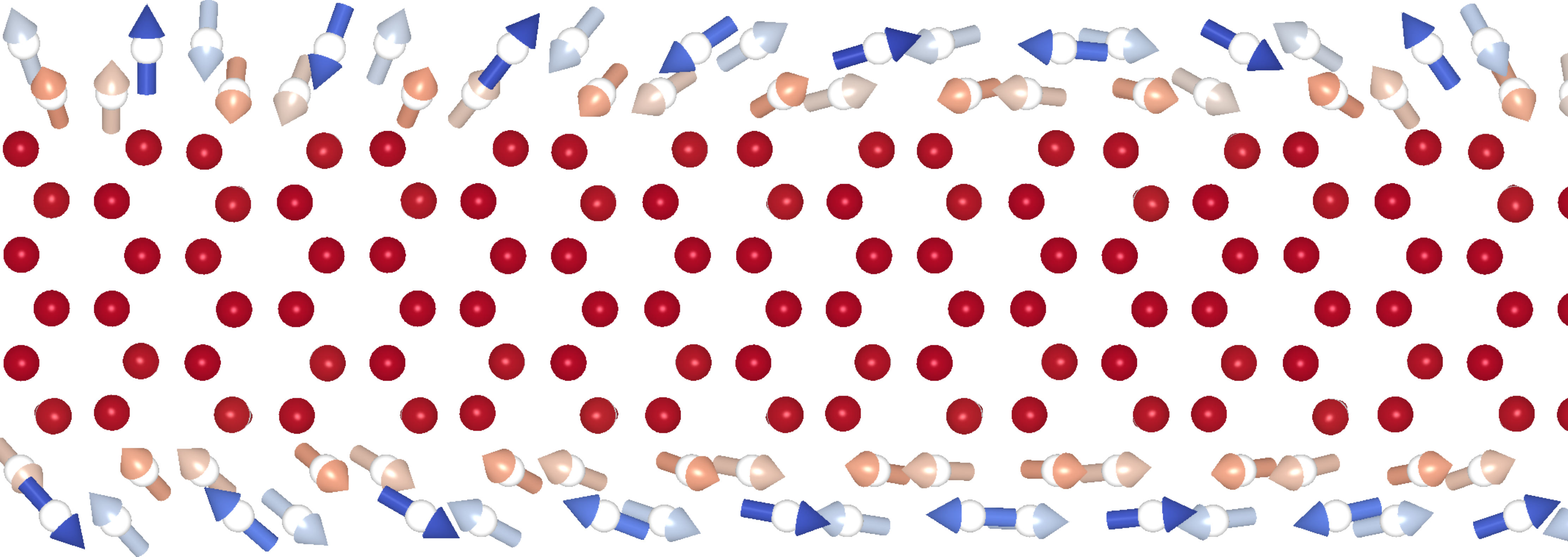} \\ \\
            \vspace{0.5cm} 
            (b) \\ \includegraphics[width=0.95\columnwidth]{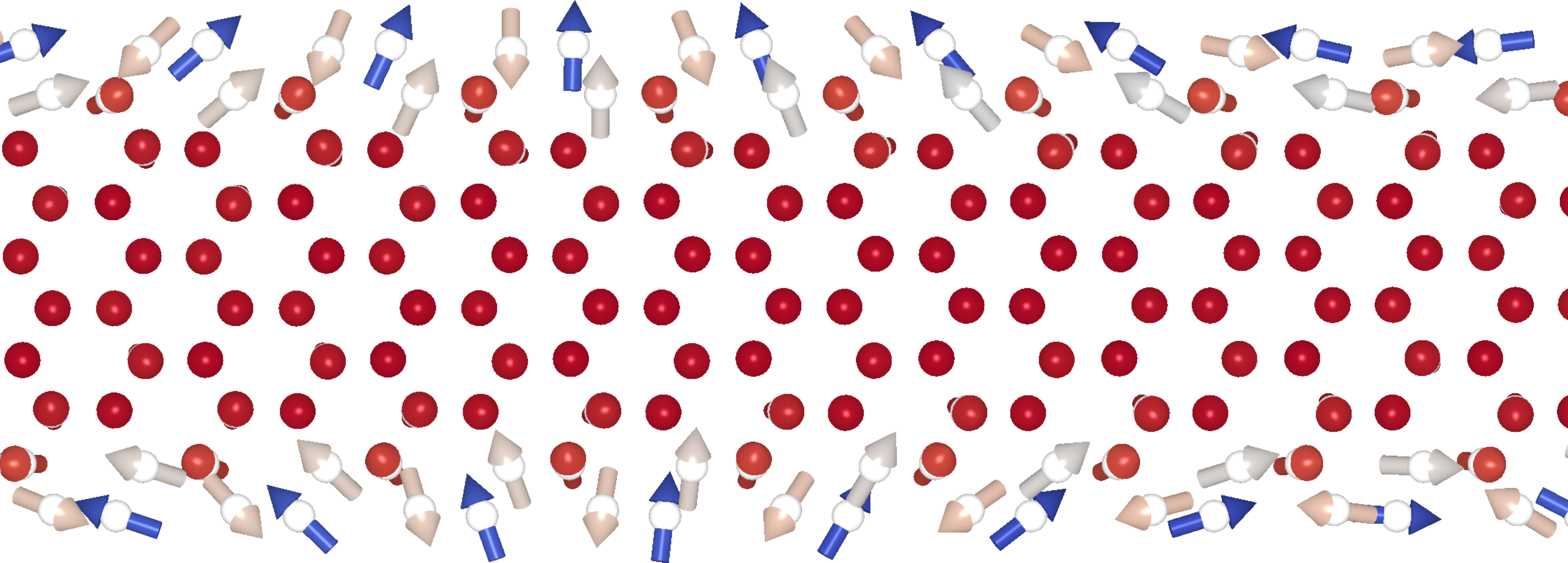}
        \end{tabular} 
    \end{tabular}
    \caption{Ground-state spin configurations of armchair-edged nanoribbons made from single-layers of FGaT (a) and FGeT (b), both with approximately 18~\AA~in breadth.}
    \label{fig:mag-gs}
\end{figure}

These parameters have been employed to determine the magnetic configuration in ground state of the nanoribbons by using the effective spin Hamiltonian given by Eq.~\ref{eq:Hamilton0}, following the methodology of the UppASD code. The results are illustrated in Fig.~\ref{fig:mag-gs}, and clearly reveal the non-collinear ground-state spin arrangements at their edges. This feature greatly expands the possibilities of manipulating the magnetization of nanostructures of these materials by means of angular-momentum torques, as they become effective regardless of the angular-momentum polarization direction. 

Moreover, when an electron hops between atoms with non-collinear magnetic moments, it generally acquires a Berry phase, which acts as a fictitious external magnetic field. This is an important ingredient in engineering $p$-wave superconductors~\cite{chatterjee_topological_2024} and also one of the mechanisms behind the existence of the spin-Hall effect in the absence of spin-orbit coupling~\cite{zhang_spin_2018}. Recently, a long-range Josephson supercurrent has been observed flowing across a flake of FGeT that connects two spin-singlet superconductors composed of NbSe$_2$~\cite{hu_long-range_2023}. A singlet spin supercurrent is predicted to decay rapidly upon entering a FM such as FGeT. However, it has been observed that the supercurrent survived for much longer distances than expected and exhibits higher density at the flake edges. Our results support the hypothesized presence of non-collinear magnetism in the FGeT flake that could promote electronic transitions to a spin-triplet state, thereby reducing the supercurrent damping, particularly at the edges.

\textit{Magnetization switching via \textup{STT} and \textup{SOT}- }
The possibility of field-free switching of magnetic states in vdW magnets is very interesting from a technological standpoint, and a recent study has indicated the possibility of current-induced switching of FGaT flakes deposited on Pt~\cite{yun_efficient_2023}. In addition, the noncollinear anti-ferromagnet Mn$_3$Sn has shown potential for efficient SOT-driven field-free switching owing to its strong magnetic spin Hall effect (mSHE)~\cite{hu_efficient_2022}. Thus, we here study how the magnetic states of FGaT and FGeT nanoribbons can be manipulated by current-induced spin torques, assuming that they are deposited on top of Mn$_3$Sn.
In the simulations, we focus on the effect of SOT with the assumption that the dominant contribution in these systems comes from the mSHE~\cite{manchon_current-induced_2019} of the Mn$_3$Sn substrate. How to incorporate these torques into an ASD framework has been explained in detail by Meo et al.~\cite{meo_spin-transfer_2022}. The strength and effectiveness of the SOT is determined by the current density $\vec{j}$, the spin-Hall angle $\theta_\text{SH}$, and the Gilbert damping parameter $\alpha$. Further details on the SOT treatment in our simulations are given in the SM. The values of $\theta_\text{SH}=0.3$ are taken from the experimental data reported in Ref.~\cite{hu_efficient_2022} to emulate the situation where the nanoribbons are deposited on Mn$_3$Sn.

\begin{figure}[htp]
    \centering
    \includegraphics[width=1.0\linewidth]{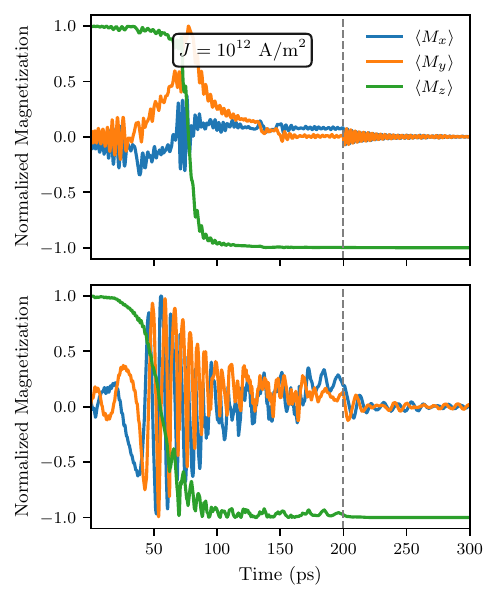}
    \caption{Average projected magnetization of the FGaT (top) and FGeT (bottom) nanoribbons during the spin dynamics process. During the first 200 ps, we show the magnetization switch dynamics via spin-orbit torque (SOT) and in the last 100 ps (after the vertical dashed line) the system is allowed to relax. The current applied is $j=10^{12}$ A/m$^2$ and $\theta_\text{SH}=0.3$.}
    \label{fig:SOT}
\end{figure}

 In Fig.~\ref{fig:SOT} we show the magnetic response of FGaT and FGeT nanoribbons when excited with a current density $j=10^{12} A/m^2$ and a polarization opposite to their initial magnetization. The temperature of the systems is fixed to T=1.0 K. For both nanoribbons we observe a switching on a time scale of less than 100 ps, showing the feasibility of rapid manipulation of these spin states by reasonable current densities. Comparing the magnetization switching processes of the two ribbons, we note more pronounced oscillation of the x- and y-components of the magnetization in FeGeT. These fast oscillations indicate that the FGeT ribbon exhibits a precession-dominated switching, while the less pronounced oscillations and the presence of plateaus in the magnetization for the FGaT ribbon suggest a switching that is dominated by magnetic domains formation. Videos illustrating the dynamics of magnetization switching, derived from our simulations for both nanoribbons, are available in the SM.

Figure \ref{fig:SOT} also illustrates what happens when the excitation current in our simulations is turned off some time after the magnetization has been switched. At $t = 200$ ps, a dashed vertical line marks this moment. From this point onwards, we observe that the expectation value of $M_z$ remains stable, while $\langle M_x\rangle$ and $\langle M_y\rangle$ gradually relax to zero.

\textit{Conclusion- }In summary, we have demonstrated that nanoribbons of Fe$_3$GeTe$_2$ and Fe$_3$GaTe$_2$ exhibit unique edge non-collinear magnetic configurations, arising from a balance between exchange and Dzyaloshinskii-Moriya interactions at the edges, which are much different from the monolayer interactions. This feature not only enables effective magnetization manipulation via spin-transfer and spin-orbit torques but also allows for rapid magnetization switching at significantly lower current densities compared to conventional FM systems. These findings emphasize the promising applications of Fe$_3$GeTe$_2$ and Fe$_3$GaTe$_2$ in spintronic, orbitronic, and magnonic devices, offering solutions for non-volatile memory and high-speed data processing. Moreover, the realization of non-collinear magnetism at the edges also sheds light in previous experiments and offers a set of new possibilities to be explored further. Our theoretical predictions could be experimentally validated using a combination of high-resolution magnetic imaging and transport techniques. For example, spin-polarized scanning tunneling microscopy (SP-STM) could directly image the predicted edge non-collinear magnetic configurations at the nanoscale. Together, our results establish these materials as versatile platforms for advancing the frontiers of low-dimensional magnetism and device miniaturization.

\textit{Acknowledgments-} R.C and R.B.M. acknowledge financial support from FAPERJ grant number E-26/205.956/2022 and 205.957/2022 (282056). R.B.M. also ackowledges the INCT of Spintronics and Advanced Magnetic Nanostructures, CNPq, Brazil. A.B. acknowledges eSSENCE and the Carl Trygger Foundation (CTS). The computations handling were enabled by resources provided by the National Academic Infrastructure for Supercomputing in Sweden (NAISS), partially funded by the Swedish
Research Council through grant agreement no. 2022-06725. M.C. acknowledges the financial support of CNPq grant No. 317320/2021-1, FAPERJ grant No.E26/200.240/2023 and INCT Materials Informatics.
\appendix

\bibliography{references} 

\end{document}